\documentclass[12pt]{article}
\usepackage{graphics}

\newcommand\scalemath[2]{\scalebox{#1}{\mbox{\ensuremath{\displaystyle #2}}}}

\newcommand{\be}{\begin{equation}}
\newcommand{\ee}{\end{equation}}
\newcommand{\bea}{\begin{eqnarray}}
\newcommand{\eea}{\end{eqnarray}}
\newcommand{\ba}{\begin{array}}
\newcommand{\ea}{\end{array}}
\begin{document}
%\begin{flushright}
%preprint number
%\end{flushright}
%
\vspace*{1.0cm}

\begin{center}
\baselineskip 20pt
{\Large\bf
Examining  A Renormalizable Supersymmetric SO(10) Model
}
\vspace{1cm}

{\large
Zhi-Yong Chen \footnote{ E-mail: chenzhiyongczy@pku.edu.cn}
and
Da-Xin Zhang\footnote{ E-mail: dxzhang@pku.edu.cn}
}
\vspace{.5cm}

{\baselineskip 20pt \it
School of Physics and State Key Laboratory of Nuclear Physics and Technology, \\
Peking University, Beijing 100871, China}

\vspace{.5cm}

\vspace{1.5cm} {\bf Abstract}
\end{center}
We examine a  renormalizable SUSY SO(10) model without fine-tuning.
We show how to construct MSSM doublets
and to predict proton decay.
We find that in the minimal set of Yukawa couplings the model is consistent with the
experiments, while including $120_H$ to fit the data there are inconsistencies.

\thispagestyle{empty}

%\bigskip
\newpage

\section{Introduction}

Supersymmetric (SUSY) Grand Unification Theories (GUTs)  of SO(10)\cite{so10a,so10b}
are very important candidates for the new physics
beyond the Standard Model (SM).
As was firstly occurred  in the SU(5) models, a very serious difficulty in all GUT models is the realization of
doublet-triplet splitting (DTS) within the same
Higgs multiplets.
The two Higgs doublets of the Minimal Supersymmetric Standard Model (MSSM)
have weak scale masses, while the color triplets and anti-triplets in the same representations
need to have masses of the GUT scale.
This is not only needed in the realization of gauge coupling unification\cite{sen1,sen2,luo1,luo2,luo3,luo4},
but  also
needed in the suppression of proton decay mediated by the colored  Higgsinos\cite{su5pd1,su5pd2,su5pd3,su5pd4}.

In the models without natural DTS, the two Higgs doublets of the MSSM
are generated through fine-tuning the doublet mass matrix\cite{so10c,so10d,so10e,so10f,fuku,so10g}.
The condition is highly nonlinear,
so that it is difficult for these doublets to satisfy those constraints got by fitting the SM fermion masses
and mixing.
Consequently, the superpotential parameters are difficult to fix so that the models
are difficult to make definite predictions on data like proton decay.

In the present work, we will discuss the MSSM doublets and proton decay in
a renormalizable SUSY SO(10) model\cite{cz2}
in which the DTS is realized through the Dimopoulos-Wilczek (DW)
mechanism\cite{dw1,dw2,dw3,dw4,lee,dw6,dw7,dw8,dw9,dw10,dw11,dw12,dw13,cz}
of missing Vacuum Expectation Value (VEV).
In this model, the MSSM doublets are linear combinations of the Higgs doublets from several
different representations of SO(10).
Consequently, the superpotential parameters are easily related to these doublets.
Then the color-triplet Higgs mass matrix is determined, which makes
the determination of proton decay feasible.
Being a renormalizable model, the $Z_2$ subgroup of the SO(10) centre $Z_4$ remains unbroken
which acts as the matter parity, thus dangerous dimension-4 operators of proton decay are eliminated,
and the lightest SUSY particle (LSP) is stable which is a good candidate of the dark matter \cite{kls}.

There are also other important advantages in the model \cite{cz2}.
Following \cite{dlz,dlz2,lz},
the type-I seesaw mechanism\cite{see1,see2,see3,see4,see5,see6,see7} for neutrino masses and mixing  is incorporated without
introducing a real scale. Instead, only a VEV smaller than the GUT scale
is used, so that all heavy particles are given masses of the GUT scale.
Consequently, the mass splitting among them is not large and thus the threshold
effects of the GUT scale can be small, at least when we adjust the parameters which are
not used in the present numerical  study. In this sense the
gauge coupling unification is maintained.
Furthermore,
the form of the color triplet mass matrix exhibits proton decay suppression explicitly\cite{dlz2,lz}.
However, whether the model \cite{cz2} is realistic or not depends on whether its prediction
on proton decay is consistent with the data, and we will examine this in the following.

\section{Superpotential}

The model in \cite{cz2} is a renormalizable SUSY SO(10) model in which the Yukawa couplings are given
by the superpotential
\be
W^{Fermion}=Y_{10}^{ij}\psi_i\psi_j H_1+Y_{120}^{ij}\psi_i\psi_j D_1+Y_{126}^{ij}\psi_i\psi_j \overline{\Delta}_1
\ee
which is generally enough to fit all fermion masses and mixing.
Here $\psi_i (i=1,2,3)$ are the matter superfields, $H_1, D_1$ and $\overline{\Delta}_1$ are the
Higgs superfields in the $10,120$ and $\overline{126}$ representations of SO(10), respectively.

SO(10) symmetry is broken by $\Phi(210)$ and/or $A(45)+E(54)$ into $SU(3)_C\times SU(2)_L\times U(1)_{B-L}\times U(1)_{I_{3R}}$. To further break $U(1)_{B-L}\times U(1)_{I_{3R}}$ into $U(1)_Y$ of the MSSM symmetry,
the SM singlets which carry nonzero $B-L$ quantum numbers are needed to have VEVs.
In renormalizable models these SM singlets are contained in $\overline{\Delta} ( \overline{126})+\Delta (126)$.
It has been studied in \cite{so10e} that these VEVs $\overline{v}=v$, which is required by the D-flatness of SUSY,
should be taken at the GUT scale $2\times 10^{16}$GeV to avoid breaking gauge coupling unification.
However, to generate low energy neutrino masses and mixing, the seesaw mechanism requires
these VEVs to be of order $\sim 10^{14}$GeV. To solve this discrepancy, we need to introduce two pairs of $\overline{\Delta}+\Delta $, one $\overline{\Delta}_1$ couples to the matter fields through (2.1) which has
a smaller VEV $\overline{v_1}\sim 10^{14}$GeV for the seesaw mechanism, the other VEVs are at the GUT scale
to realize gauge coupling unification.

To be specific, we introduce the following superpotential
\be
%W_{D\Delta}^0= +
 \left(m_1+\eta_{1}\Phi  \right) \overline{\Delta_1}\Delta_2
+\left(m_2+\eta_{2}\Phi  \right) \overline{\Delta_2}\Delta_1+\eta_{3}Q\overline{\Delta }_2\Delta_2\label{W0}
\ee
which contains all interactions relevant to the  $U(1)_{B-L}$ breaking.
The D-flatness condition of maintaining SUSY at the GUT scale is simply
$|\overline{v_1}|^2+|\overline{v_2}|^2=|{v_1}|^2+|{v_2}|^2$.
The F-flatness conditions are
  \bea 0&=&
  \left(\ba{cc}\overline{v}_{1} & \overline{v}_{2}\ea\right)
  \left(\ba{cc} 0 & m_1 +\eta_{1}\Phi_0\\m_2 +\eta_{2}\Phi_0&\eta_{3}Q \ea\right), \nonumber\\
  0&=&\left(\ba{cc} 0 & m_1 +\eta_{1}\Phi_0\\m_2 +\eta_{2}\Phi_0&\eta_{3}Q\ea\right)
  \left(\ba{c} {v}_{1} \\ {v}_{2} \ea \right),\nonumber
  \eea
where $\Phi_0$ is a combination of  the three VEVs in $\Phi$, and the SO(10) singlet $Q$ has a VEV
$\sim 10^{-2}\Lambda_{GUT}$. This smaller VEV $Q$ can be linked with the Planck scale by
$Q\sim \Lambda_{GUT}^2/\Lambda_{Planck}$
through the Green-Schwarz mechanism\cite{green1,green2,green3,green4},
provided that the extra global symmetry is embedded into an anomalous U(1) symmetry\cite{lz}.
One set of solutions of the above equations require $m_2 +\eta_{2}\Phi_0=0$, which gives
\begin{equation}
\overline{v}_{1} = \overline{v}_{2} \frac{\eta_{3}Q}{\frac{\eta_{1}}{\eta_{2}}m_{2} - m_{1}}\sim 10^{-2}\Lambda_{GUT}
\label{v1}
\end{equation}
and
\be
v_2=0,\label{v20}
\ee
and $\overline{v}_{2}\sim v_1\sim \Lambda_{GUT}$ satisfying the D-flatness condition.
Note that following (\ref{W0}),
the color triplet-anti-triplet mass matrix has the structure
\be
\left(
\ba{cc} 0& \Lambda_{GUT}\\ \Lambda_{GUT}& Q
\ea
\right).\label{126mass}
\ee
When only $\overline{\Delta_1}$ couples with the matter superfields, we can integrate out $\Delta_2-\overline{\Delta }_2$
to generate the effective triplet mass matrix whose elements are $\sim \frac{ \Lambda_{GUT}^2}{Q}\sim 10^2 \Lambda_{GUT}$.
Consequently, the dimension-five operators for proton decay mediated by the color triplet-anti-triplet of
$\overline{\Delta_1}$ are suppressed accordingly. To suppress proton decay mediated by $D_1(120)$ of (2.1), we need also
introduce a second $D_2(120)$ to get a mass matrix with the same structure as (\ref{126mass})..

Now we include $H_1(10)$ which couples with the matter superfields.
$H_1,D_1,\Delta_1,\overline{\Delta }_1$  couple with $D_2,\Delta_2,\overline{\Delta }_2$ through $\Phi$.
Note that in $\Phi$ there is also a pair of $SU(2)_L$ doublets.
As will be seen in the next Section,
${v}_{2}=0$ eliminates some possible mixing terms between the doublets of
$\Phi$ and those of $H_1,D_1, {\Delta}_1,\overline{\Delta}_1$.
This is crucial in generating a pair of massless doublets in the model.
There are, however, also a pair of  massless color triplet-anti-triplet.
To give these triplet masses of the GUT scale, we need to use the DW mechanism using
a second $H_2(10)$ and $A(45)$ with DW-type VEVs $A_1=0, A_2\neq 0$.
Here $A_1$ and $A_2$ are the VEVs of the SM singlets
in the (1,1,3) and (15,1,1) directions, respectively,
under  the $SU(4)_c\times SU(2)_L\times SU(2)_R$ subgroup of SO(10).
To suppress $H_1(10)$ mediated dimension-five operators for proton decay,
we need another $A'(45)$ whose VEVs  $A'_1\neq 0, A'_2=0$ which are the compliment to the DW (CDW) mechanism.
In \cite{cz2} it was found that the simplest method to realize both the DW and CDW mechanisms
is using the superpotential of the form
\be
\xi_{1}P A A'+(\xi_{2}E+\xi_{3}R) A' A''
\ee
which contains all interactions for $A'$ and the new $A''(45)$ contributing to their F-terms.
Here $P,R$ are SO(10) singlets playing the roles of masses and $E$ is a $54$ of SO(10).
Solving the F-term conditions for $A''$ gives the CDW VEV $A'_2=0$,
then solving the F-term conditions for $A'$ gives the DW VEV $A_1=0$.
There is another set of solutions with $A'_1=0$ and $A_2=0$ which are not used.

In the renormalizable models, the direct application of the DW mechanism does not work,
since the couplings $H_1 A H_2$ and $D_1 A H_2$ exist simultaneously. The later interaction
invalidities  the DW mechanism due to the coupling $D_1(15,2,2) A_2(15,1,1) H_2(1,2,2)$
using the notations under $SU(4)_c\times SU(2)_L\times SU(2)_R$.
To avoid this interaction, the  filter mechanism\cite{cz} can be used with the superpotential
\be
P H_1 \overline{h}+ m_h h \overline{h}+A h H_2+\frac{1}{2}m_{H_2} H_2^2\label{dw0}
\ee
where dimensionless couplings are suppressed.
The singlet $P$ is used as a  filter to eliminate  $ D_1 P A H_2$ while keeping $ H_1 P A H_2$.
To apply the CDW mechanism to suppress proton decay mediated by the color triplets of $H_1$,
the last term in (\ref{dw0}) is replaced by
\be
A' H_2 H_3 + \frac{1}{2}m_{H_3}H_3^2.
\ee
Consequently, $H_1$ mediated proton decay is fobidden, as can be seen in Section 3.
In building realistic models, mass parameters can be replaced by VEVs of singlets and/or $54$s.

When we use all the above superfields to build the model, the F-flatness conditions cannot be
all consistent so that a new $E'(54)$ is introduced.
All the superfields  are summarized in Table 1.
To avoid unwanted terms,
we have enforced an extra symmetry $Z_{24}\times Z_4$ under which the transformation properties of all the
particles are also listed in Table 1.
Note that to generate the seesaw VEV $\sim Q$ through the Green-Schwarz mechanism,
this discrete symmetry is the subgroup of the anomalous gauge U(1) groups \cite{lz},
its symmetry breaking may not bring in the domain wall problem.
\begin{table}\begin{center}
\begin{tabular}{|c||c|c|c|c|c|c|c|c|c|c|c|}
  \hline  \hline
    &$A^{\prime\prime}$ & $E$ & $R$ & $A^{\prime}$ & $P$ & $A$ & $E^{\prime}$ &$\Phi$& $Q$ & $\psi_i$ &\\\hline
  SO(10) &45&54&1&45&1&45&54&210&1&16& \\\hline
  $Z_{24}$ &2 & 12 & 12 & 10 & 2 & 12 & 0 & 0 &4&-1&\\\hline
  $Z_4$&-1&0&0&1&1&2&0&0&0&0&\\\hline\hline
      & $H_1$& $\overline{h}$ & $h$& $H_2$&$H_3$ & $D_1$ &$\overline{\Delta}_1$ & ${\Delta}_1$ &$D_2$ &$\overline{\Delta}_2$ & ${\Delta}_2$ \\
  \hline
  SO(10)&10&10&10&10&10&120&$\overline{126}$&126&120&$\overline{126}$&126\\
  \hline
  $Z_{24}$ & 2&-4&4&8&6&2&2&2&-2&-2&-2\\\hline
    $Z_{4}$ &0&-1&1&1&2&0&0&0&0&0&0\\
  \hline  \hline
\end{tabular}
\caption{Notations and $Z_{24}\times Z_4$ properties of all superfields. Here $\psi_i (i=1,2,3)$ are the matter superfields.}\end{center}
\end{table}
Then the full superpotential
is
\be
W^{Higgs}=W_{SB}+W_{D\Delta}+W_{filter}+W_{DW},\label{Wfull}
\ee
where
\bea
W_{SB}&=&\frac{1}{2}m_{\Phi }\Phi ^2 +\lambda'_{1}\Phi ^3+\lambda'_{2}E^\prime \Phi^2+\lambda'_{3}\Phi A^2
                +\frac{1}{2}m_{{E'}} {E'}^2+\frac{1}{2}m_EE^2\nonumber\\
             & &+\lambda'_{4}E^2 {E'}+\lambda'_{5}{E'}^3+\frac{1}{2}m_AA^2+\lambda'_{6}{E'}A^2 +\frac{1}{2} m_R R^2
                +\lambda'_{7}R E E^\prime,\nonumber\\
W_{D\Delta}&=& \left(\eta'_{1}\Phi +m_1 \right) \overline{\Delta_1}\Delta_2
+\left(\eta'_{2}\Phi +m_2 \right) \overline{\Delta_2}\Delta_1+\eta'_{3}Q\Delta_2\overline{\Delta }_2\nonumber\\
& &+\Phi D_1 \left(\eta'_{4}\Delta_2+\eta'_{5}\overline{\Delta_2} \right)
+\Phi \left(\eta'_{6}\Delta_1+ \eta'_{7}\overline{\Delta_1} \right) D_2+\eta'_{8}Q D_2^2\nonumber\\
& &+ E^\prime (\eta'_{9}\Delta_1\Delta_2+\eta'_{10}\overline{\Delta_1}\overline{\Delta_2})\nonumber\\
& &+ \Phi H_1\left(\eta'_{11}D_2+\eta'_{12}\Delta_2+\eta'_{13}\overline{\Delta_2} \right)
+\left(m_D+\eta'_{14}{E'}+\eta'_{15}\Phi\right) D_1 D_2,\nonumber\\
W_{filter}&=&\kappa_{1}P H_1 \overline{h}+ \left(\kappa_{2}{E'}+m_h\right)h \overline{h}+\kappa_{3}A h H_2+\kappa_{4}A' H_2 H_3 \nonumber\\
                 & &+ \frac{1}{2}(\kappa_{5}R+\kappa_{6}E)H_3^2.\nonumber\\
W_{DW}&=&\xi_{1}P A A'+(\xi_{2}E+\xi_{3}R) A' A''.\nonumber
\eea

Here, the couplings with ``$^\prime$'' follow the notations given in \cite{so10f}. However,
not all of them are normalized properly to be of order one.
In Table 2, we redefine these couplings so that the unprimed couplings are of order one
numerically.
\begin{table}\begin{center}
		\begin{tabular}{|c||c|c|c|c|c|c|c|c|c|}
			\hline  \hline
			Old & $\eta'_1$ &$\eta'_2$ & $\eta'_3$ &  $\eta'_4$   & $\eta'_5$ &$\eta'_6$ & $\eta'_7$ &$\eta'_8$&$\eta'_9$ \\ \hline
			New & $10\sqrt{6}\eta_1$ & $10\sqrt{6}\eta_2$ & $\eta_3$  &$2\sqrt{30}\eta_4$& $2\sqrt{30}\eta_5$   & $2\sqrt{30}\eta_6$&$2\sqrt{30}\eta_7$ & $\eta_8$ & $\frac{5\sqrt{2}}{2}\eta_9$   \\
			\hline
			Old &$\eta'_{10}$ &$\eta'_{11}$ & $\eta'_{12}$& $\eta'_{13}$& $\eta'_{14}$& $\eta'_{15}$ &$\lambda'_{1}$ & $\lambda'_{2}$&  \\ \hline
			New &  $\frac{5\sqrt{2}}{2}\eta_{10}$ &$2\eta_{11}$ & $\sqrt{5}\eta_{12}$ & $\sqrt{5}\eta_{13}$ & $\frac{3\sqrt{2}}{2}\eta_{14}$ &  $\frac{3\sqrt{6}}{2}\eta_{15}$ & $\sqrt{6}\lambda_{1}$ & $2\sqrt{2}\lambda_{2}$& \\
            \hline \hline
		\end{tabular}
        \caption{Redefinitions of the couplings.}
	\end{center}
\end{table}

Compared to \cite{cz2}, we have eliminated a reluctant $45$ and its interactions.
Although the superpotential (\ref{Wfull}) is complicated, it solves several major
difficulties of SUSY SO(10) at the same time and thus can be taken as a prototype of realistic SUSY GUTs.
The many representations used in building this model may bring in the question  if they are
allowed. To our knowledge, except  the string model argument \cite{dienes}
based on perturbative study, there is no reason to exclude these representations  in principle.
Whether the model is minimal or not remains an open question.

\section{The weak doublets and the color triplets}

The mass matrix for the doublets can be read off from (\ref{Wfull}).
To simplify the discussion,
we neglect  $W_{filter}$ at first and consider its effects later.
The mass matrix for the doublets is
\begin{eqnarray}\label{massDf}
M_D^{D\Delta}=\left(
\begin{array}{c|c}
0_{6\times 5}&A_{6\times 5}\\
\hline
B_{4\times 5}&C_{4\times 5}
\end{array}
\right),
\end{eqnarray}
where the columns are
($H_1^u,D_1^u,D_1^{\prime u},\overline{\Delta}^u_1,{\Delta}^u_1;
{\Phi}^u;\overline{\Delta}^u_2,{\Delta}^u_2 ,D_2^u,D_2^{\prime u}$),
and the rows are
 \\
 ($H_1^d,D_1^d,D_1^{\prime d},{\Delta}^d_1,\overline{\Delta}^d_1;
{\Phi}^d;{\Delta}^d_2 ,\overline{\Delta}^d_2,D_2^d,D_2^{\prime d}$).
The 6th row corresponds to $\Phi^d$, and the first 5 entries in this row are
proportional to $v_{2}$ which is zero.
It is obvious that the upper-most 6 rows are not independent
which combine into a massless eigenstate of $H_d$ type,
while the left-most 5 columns give a $H_u$ type massless eigenstate.
Consequently, the massless doublets can be easily seen from (\ref{massDf}),
\bea
H_{u}^{0}&=&\alpha_{u}^{1}H_{1}^{u}+\alpha_{u}^{2}D_{1}^{u}+\alpha_{u}^{3}D_{1}^{\prime u}
+\alpha_{u}^{4}\overline{\Delta}_{1}^{u}+\alpha_{u}^{5}\Delta_{1}^{u}\nonumber\\
H_{d}^{0}&=&\alpha_{d}^{1}H_{1}^{d}+\alpha_{d}^{2}D_{1}^{d}+\alpha_{d}^{3}D_{1}^{\prime d}
+\alpha_{d}^{4}\Delta_{1}^{d}+\alpha_{d}^{5}\overline{\Delta}_{1}^{d}+\alpha_{d}^{6}\Phi_{1}^{d}\label{H0}
\eea
satisfying the linear equations
\bea
(\alpha_{d}^{1},\alpha_{d}^{2},\alpha_{d}^{3},\alpha_{d}^{4},\alpha_{d}^{5},\alpha_{d}^{6})^{*}
A_{6\times5}=0, ~~
B_{4\times5}
(\alpha_{u}^{1},\alpha_{u}^{2},\alpha_{u}^{3},\alpha_{u}^{4},\alpha_{u}^{5})^{\dagger}=0\label{eigen}
\eea
and the normalization conditions
\be
1=|\alpha_{u}^{1}|^2+|\alpha_{u}^{2}|^2+|\alpha_{u}^{3}|^2+|\alpha_{u}^{4}|^2+|\alpha_{u}^{5}|^2,
~~~1=|\alpha_{d}^{1}|^2+|\alpha_{d}^{2}|^2+|\alpha_{d}^{3}|^2+|\alpha_{d}^{4}|^2+|\alpha_{d}^{5}|^2
+|\alpha_{d}^{6}|^2.
\ee

The explicit forms of $A_{6\times5}$ and  $B_{4\times5}$ can be read off from the superpotential,

\begin{eqnarray}
A_{6\times5} =
\left(
\begin{array}{ccccc}
 -\eta _{13} \bar{v}_2 &
 a_{12} &
 \frac{1}{2} \eta _{12} \left(\sqrt{2} \Phi _2-\Phi _3\right)&
 -\eta _{11} \Phi _1 & -\frac{\eta _{11} \Phi _3}{\sqrt{2}} \\
 -\eta _5 \bar{v}_2 &
 \frac{\eta _5 \Phi _3}{2} &
 \frac{\eta _4 \Phi _3}{2} & a_{24} & \frac{\eta _{15}\Phi _3}{2 \sqrt{2}} \\
 -\sqrt{3} \eta _5 \bar{v}_2 &
  a_{32} &
   \frac{1}{6} \eta _4 \left(3 \sqrt{2} \Phi_1-2 \sqrt{3} \Phi _3\right) &
   \frac{\eta _{15} \Phi _3}{2 \sqrt{2}} &
    m_D-\frac{E' \eta _{14}}{2 \sqrt{30}}+\frac{\eta _{15} \Phi _2}{\sqrt{3}}\\
 \sqrt{6} \eta _2 \bar{v}_2 &
  a_{42} &
  \frac{1}{2} \sqrt{\frac{5}{6}} E' \eta _9 &
\frac{\eta _6 \Phi _3}{2} &
\eta _6 \left(\frac{\Phi _1}{\sqrt{2}}+\frac{\Phi _3}{\sqrt{3}}\right) \\
 0 &
  \frac{1}{2} \sqrt{\frac{5}{6}} E' \eta _{10} &
   m_1+\frac{\eta _1 \left(2 \Phi _2-\sqrt{2} \Phi _3\right)}{\sqrt{3}} &
   \frac{\eta _7 \Phi _3}{2}&
   \frac{1}{6} \eta _7 \left(3 \sqrt{2} \Phi _1-2 \sqrt{3} \Phi _3\right) \\
a_{61}&
\sqrt{6} v_1 \eta _2 &
 0 &
  -v_1\eta _6 &
  -\sqrt{3} v_1 \eta _6 \\
\end{array}
\right), \nonumber
\end{eqnarray}
where
\begin{eqnarray}
a_{12} &=& -\frac{1}{2} \eta _{13} \left(\sqrt{2} \Phi _2+\Phi _3\right),\nonumber \\
a_{24} &=& m_D+\frac{3}{2} \sqrt{\frac{3}{10}} E' \eta _{14}, \nonumber \\
a_{32} &=& \eta_5 \left(\frac{\Phi _1}{\sqrt{2}}+\frac{\Phi _3}{\sqrt{3}}\right), \nonumber \\
a_{42} &=& m_2+\frac{\eta _2 \left(2 \Phi _2+\sqrt{2} \Phi _3\right)}{\sqrt{3}},\nonumber \\
a_{61} &=&  m_{\Phi }-\sqrt{\frac{3}{10}} E' \lambda _2+\sqrt{3} \lambda _1 \Phi _2+\sqrt{\frac{3}{2}} \lambda _1 \Phi _3, \nonumber
\end{eqnarray}
and
\begin{eqnarray}
B_{4\times5} =
\left(
\begin{array}{ccccc}
  - \eta _{12} \left(\frac{\sqrt{2}}{2} \Phi _2+\frac{1}{2}\Phi _3\right) &
  \frac{\eta _4 \Phi _3}{2} &
  \eta _4 \left(\frac{\Phi _1}{\sqrt{2}}+\frac{\Phi _3}{\sqrt{3}}\right)&
  b_{14}&
  \frac{1}{2} \sqrt{\frac{5}{6}} E' \eta _9 \\
  \frac{1}{2} \eta _{13} \left(\sqrt{2} \Phi _2-\Phi _3\right) &
  \frac{\eta _5 \Phi _3}{2} &
  b_{23}&
  \frac{1}{2} \sqrt{\frac{5}{6}} E' \eta _{10} &
  b_{25} \\
  -\eta _{11} \Phi _1 &
  b_{32}  &
  \frac{\eta _{15} \Phi _3}{2 \sqrt{2}} &
  \frac{\eta _7 \Phi _3}{2} &
  \frac{\eta_6 \Phi _3}{2} \\
  -\frac{\eta _{11} \Phi _3}{\sqrt{2}} &
  \frac{\eta _{15} \Phi _3}{2 \sqrt{2}} &
  b_{43}&
  \eta _7 \left(\frac{\Phi _1}{\sqrt{2}}+\frac{\Phi _3}{\sqrt{3}}\right) &
  b_{45}

\\
\end{array}
\right), \nonumber
\end{eqnarray}
where
\begin{eqnarray}
b_{14} &=& m_1+\frac{\eta _1 \left(2 \Phi _2+\sqrt{2} \Phi _3\right)}{\sqrt{3}},  \nonumber \\
b_{23} &=& \eta _5 \left( \frac{\sqrt{2}}{2} \Phi _1 - \frac{\sqrt{3}}{3}  \Phi _3\right), \nonumber \\
b_{25} &=& m_2+\frac{\eta _2 \left(2 \Phi _2-\sqrt{2} \Phi _3\right)}{\sqrt{3}}, \nonumber \\
b_{32} &=& m_D+\frac{3}{2} \sqrt{\frac{3}{10}} E' \eta _{14}, \nonumber \\
b_{43} &=& m_D-\frac{E' \eta _{14}}{2 \sqrt{30}}+\frac{\eta _{15} \Phi _2}{\sqrt{3}}, \nonumber \\
b_{45} &=& \eta _6 \left( \frac{\sqrt{2}}{2} \Phi _1 - \frac{\sqrt{3}}{3}  \Phi _3\right). \nonumber
\end{eqnarray}

Here we have used the fields to represent the VEVs of their SM singlets without introducing
confusion, and $\Phi_{1,2,3}$ are the VEVs of $\Phi$ in the (1,1,1), (15,1,1) and (15,1,3) directions,
respectively, under the $SU(4)_c\times SU(2)_L\times SU(2)_R$ subgroup.

Now we take into the effects of $W_{filter}$. Ordering both the columns and the rows as
$(H_{u(d)}^0, \overline{h}, h, H_2,H_3$)
with $H_{u,d}^0$ given in (\ref{H0}), we have the doublet mass matrix
\begin{eqnarray}\label{filterMD}
M_D^{filter}=\left(
\begin{array}{ccccc}
0 & \alpha_d^1 \kappa_{1}P & 0 & 0 &0\\
\alpha_u^1 \kappa_{1}P & 0 &\kappa_{2}{E'}+m_h& 0 &0\\
0 &\kappa_{2}{E'}+m_h& 0 &\kappa_{3}A_1=0&0\\
0 & 0 &\kappa_{3}A_1=0&0&\kappa_{4}A_1^\prime\\
0&0&0&\kappa_{4}A_1^\prime&\kappa_{5}R+\kappa_{6}E
\end{array}
\right),
\end{eqnarray}
then we have a pair of massless eigenstates
\begin{eqnarray}\label{doub2}
H_u&=&\frac{(\kappa_{2}{E'}+m_h) H_u^0-(\alpha_u^1 \kappa_{1}P) h^u}{\sqrt{|\alpha_u^1 \kappa_{1}P|^2
+|\kappa_{2}{E'}+m_h|^2}}, \nonumber\\
H_d&=&\frac{(\kappa_{2}{E'}+m_h) H_d^0-(\alpha_d^1 \kappa_{1}P) h^d}{\sqrt{|\alpha_d^1 \kappa_{1}P|^2
+|\kappa_{2}{E'}+m_h|^2}},
\end{eqnarray}
which  are the weak doublets in the MSSM.
For $P$ having a VEV of order $\Lambda_{GUT}$
the components of $H_1^{u,d}$ in the MSSM doublet $H_{u,d}$ are not small,
thus we can take (\ref{doub2}) as pure normalization without significant numerical effects,
and we will neglect these effects in the doublets to simplify our discussions.

In the absence of the effects from $W_{filter}$,
there are two more pairs of color triplets from $\overline{\Delta_{1,2}}+\Delta_{1,2}$ comparing to the doublets,
and the mass matrix for the triplets without the effects of $W_{filter}$ is
\begin{eqnarray}\label{massTf}
M_T^{D\Delta}=\left(
\begin{array}{c|c}
0_{7\times 6}&A_{7\times 6}\\
\hline
B_{5\times 6}&C_{5\times 6}
\end{array}
\right),
\end{eqnarray}
where the columns are
($H_1^T,D_1^T,D_1^{\prime T},\overline{\Delta}^T_1,\overline{\Delta}^{\prime T}_1,{\Delta}^T_1;
{\Phi}^T;\overline{\Delta}^T_2,\overline{\Delta}^{\prime T}_2,{\Delta}^T_2 ,D_2^T,D_2^{\prime T}$),
while the rows are similarly ordered.
Again, there is a pair of massless triplets.
We can re-write the mass matrix in (\ref{massTf}) as
\begin{eqnarray}\label{massTf2}
M_T^{D\Delta}=\left(
\begin{array}{c|c}
0_{6\times 6}&A^\prime_{6\times 6}\\
\hline
B^\prime_{6\times 6}&C^\prime_{6\times 6}
\end{array}
\right),
\end{eqnarray}
so that only the upper-left sub-matrix may couple to the matter fields.
The explicit forms of $A^{\prime}_{6\times6}$, $B^{\prime}_{6\times6}$ and $C^{\prime}_{6\times6}$ are

\begin{eqnarray}
A'_{6\times6} = \left(
\begin{array}{cccccc}
 \eta _{13} \bar{v}_2 & a_{12} & -\sqrt{\frac{2}{3}} \eta _{13} \Phi _3 & a_{14} & -\frac{\eta _{11} \Phi _3}{\sqrt{3}} & -\sqrt{\frac{2}{3}} \eta _{11} \Phi _2 \\
 -\sqrt{2} \eta _5 \bar{v}_2 & \frac{\eta _5 \Phi _3}{\sqrt{6}} & \sqrt{\frac{2}{3}} \eta _5 \Phi _2 & \frac{\eta _4 \Phi _3}{\sqrt{6}} & a_{25} & \frac{\eta _{15} \Phi _3}{\sqrt{6}} \\
 \sqrt{2} \eta _5 \bar{v}_2 & \frac{\eta _5 \Phi _2}{\sqrt{3}} & \frac{\eta _5 \Phi _3}{\sqrt{3}} & -\frac{\eta _4 \Phi _2}{\sqrt{3}} & \frac{\eta
_{15} \Phi _3}{\sqrt{6}} & a_{36} \\
 -\sqrt{2} \eta _2 \bar{v}_2 & m_2 & \frac{2 \eta _2 \Phi _3}{\sqrt{3}} & \sqrt{\frac{5}{6}} E' \eta _9 & \frac{\eta _6 \Phi _3}{\sqrt{6}} & \frac{\eta
_6 \Phi _2}{\sqrt{3}} \\
 -2 \eta _2 \bar{v}_2 & \frac{2 \eta _2 \Phi _3}{\sqrt{3}} & a_{53} & 0 & \sqrt{\frac{2}{3}} \eta
_6 \Phi _2 & \frac{\eta _6 \Phi _3}{\sqrt{3}} \\
 0 & \sqrt{\frac{5}{6}} E' \eta _{10} & 0 & m_1 & \frac{\eta _7 \Phi _3}{\sqrt{6}} & -\frac{\eta _7 \Phi _2}{\sqrt{3}} \\
\end{array}
\right)\nonumber,
\end{eqnarray}

where
\begin{eqnarray}
a_{12} &=& -\frac{\eta _{13} \left(3 \Phi _1+\sqrt{3} \Phi _2\right)}{3 \sqrt{2}}, \nonumber \\
a_{14} &=& \frac{\eta_{12} \left(-3 \Phi _1+\sqrt{3} \Phi _2\right)}{3 \sqrt{2}}, \nonumber \\
a_{25} &=& m_D+\sqrt{\frac{2}{15}}E' \eta _{14}+\frac{\eta _{15} \Phi _1}{2}, \nonumber \\
a_{36} &=& m_D-\sqrt{\frac{3}{10}} E' \eta _{14}+\frac{\eta _{15} \Phi _2}{2 \sqrt{3}}, \nonumber \\
a_{53} &=& m_2+\eta _2 \left(\Phi _1+\frac{\Phi _2}{\sqrt{3}}\right),\nonumber
\end{eqnarray}

\begin{eqnarray}
B'_{6\times6} = \left(
\begin{array}{cccccc}
 0 & 0 & 0 & 0 & 0 & 0 \\
 -\frac{\eta _{12} \left(3 \Phi _1+\sqrt{3} \Phi _2\right)}{3 \sqrt{2}} & \frac{\eta _4 \Phi _3}{\sqrt{6}} & \frac{\eta _4 \Phi _2}{\sqrt{3}} & m_1
& \frac{2 \eta _1 \Phi _3}{\sqrt{3}} & \sqrt{\frac{5}{6}} E' \eta _9 \\
 -\sqrt{\frac{2}{3}} \eta _{12} \Phi _3 & \sqrt{\frac{2}{3}} \eta _4 \Phi _2 & \frac{\eta _4 \Phi _3}{\sqrt{3}} & \frac{2 \eta _1 \Phi _3}{\sqrt{3}}
& b_{35} & 0 \\
 \frac{\eta _{13} \left(-3 \Phi _1+\sqrt{3} \Phi _2\right)}{3 \sqrt{2}} & \frac{\eta _5 \Phi _3}{\sqrt{6}} & -\frac{\eta _5 \Phi _2}{\sqrt{3}} &
\sqrt{\frac{5}{6}} E' \eta _{10} & 0 & m_2 \\
 -\frac{\eta _{11} \Phi _2}{\sqrt{3}} & b_{52} & \frac{\eta _{15} \Phi _3}{\sqrt{6}}
& \frac{\eta _7 \Phi _3}{\sqrt{6}} & \sqrt{\frac{2}{3}} \eta _7 \Phi _2 & \frac{\eta _6 \Phi _3}{\sqrt{6}} \\
 -\sqrt{\frac{2}{3}} \eta _{11} \Phi _2 & \frac{\eta _{15} \Phi _3}{\sqrt{6}} & b_{63}& \frac{\eta _7 \Phi _2}{\sqrt{3}} & \frac{\eta _7 \Phi _3}{\sqrt{3}} & -\frac{\eta _6 \Phi _2}{\sqrt{3}} \\
\end{array}
\right)\nonumber,
\end{eqnarray}

where
\begin{eqnarray}
b_{35} &=& m_1+\eta _1 \left(\Phi _1+\frac{\Phi _2}{\sqrt{3}}\right), \nonumber \\
b_{52} &=& m_D+\sqrt{\frac{2}{15}} E' \eta _{14}+\frac{\eta _{15} \Phi _1}{2}, \nonumber \\
b_{63} &=& m_D-\sqrt{\frac{3}{10}} E' \eta _{14}+\frac{\eta _{15} \Phi _2}{2\sqrt{3}}, \nonumber
\end{eqnarray}
and
\begin{eqnarray}
C'_{6\times 6} = \left(
\begin{array}{cccccc}
c_{11} & -\sqrt{2} v_1 \eta_2 & -2 v_1 \eta _2 & 0 & -\sqrt{2} v_1 \eta _6 & \sqrt{2} v_1 \eta _6 \\
 -\sqrt{2} \eta _1 \bar{v}_1 & Q \eta _3 & 0 & 0 & 0 & 0 \\
 -2 \eta _1 \bar{v}_1 & 0 & Q \eta _3 & 0 & 0 & 0 \\
 0 & 0 & 0 & Q \eta _3 & 0 & 0 \\
 -\sqrt{2} \eta _7 \bar{v}_1 & 0 & 0 & 0 & Q \eta _8 & 0 \\
 \sqrt{2} \eta _7 \bar{v}_1 & 0 & 0 & 0 & 0 & Q \eta _8 \\
\end{array}
\right),\nonumber
\end{eqnarray}

where
\begin{eqnarray}
c_{11} =  m_{\Phi }+\sqrt{\frac{2}{15}} E' \lambda _2+\frac{1}{3} \lambda _1 \left(3 \Phi _1+\sqrt{3} \Phi _2+2 \sqrt{6} \Phi _3\right). \nonumber
\end{eqnarray}

In calculating proton decay rates mediated by the color triplet Higgsinos,
what is relevant is the effective mass matrix $M^{Eff}$ which is got from the full mass matrix
by integrating out those fields which do not couple
with the matter fields.
This effective mass matrix is
$M^{Eff}=A^\prime (C^\prime)^{-1} B^\prime$.
Proton decay amplitudes depend on the inverses of the eigenvalues of $M^{Eff}$
so that small eigenvalues of $C^\prime$ are needed to suppress proton decay.
$C^{\prime}_{6\times6}$ contains 5 small eigenvalues which are not enough to generate
6 large eigenvalues for the effective mass matrix.
This is cured by including the effects from $W_{filter}$ which gives
\begin{eqnarray}\label{filterMT}
M_T^{filter}=\left(
\begin{array}{ccccc}
0 & \kappa_{1}P & 0 & 0 &0\\
\kappa_{1}P & 0 &\kappa_{2}{E'}+m_h& 0 &0\\
0 &\kappa_{2}{E'}+m_h& 0 &\kappa_{3}A_2&0\\
0 & 0 &\kappa_{3}A_2&0&\kappa_{4}A_2^\prime=0\\
0&0&0&\kappa_{4}A_2^\prime=0&\kappa_{5}R+\kappa_{6}E
\end{array}
\right),
\end{eqnarray}
where the bases are  $(H_1^T, \overline{h}^T, h^T, H_2^T,H_3^T$) in the columns
and similar for the anti-triplets in the rows.
After integrating out the fields $\overline{h}^T, h^T, H_2^T,H_3^T$ and their conjugates,
the effect is to give an infinite effective mass to $H_1^T H_1^{\overline{T}}$,
or the (1,1) entry in
$M_T^{D\Delta}$ of (\ref{massTf2}) is replaced by an infinity.
Now there are 6 large eigenvalues including an infinity in the effective triplet mass matrix which are
supposed to be
sufficient to suppress all proton decay amplitudes.

\section{Determination of the parameters}

In order to calculate the proton decay rates, we need to know the
parameters in the color triplet mass matrix. These parameters also appear in the
weak doublet mass matrix which gives the two massless doublets of the MSSM
and hence are linked to the matter masses and mixing.
There are also constraints from the neutrino oscillations\cite{fuku,fits1,fits2,yang,fits3,fits4,patel1,patel2,patel3,patel4,fits5,fits6,fits7,fits8}.
In the literature, however, since few people believe that the MSSM doublets are
got through fine-tuning the doublet mass matrix, proton decays are calculated
by simply adjusting parameters in the color-triplet mass matrix\cite{fuku,pdso10,fuku16,mo1,mo2,mo3}.
The fitting of the fermion masses and mixing can give constraints on the
components of the weak doublets, but they are not linked to the color-triplets
in the absence of a realistic mechanism of naturally generating the weak doublets.

In the model of \cite{cz2},  the doublets are got without
fine-tuning so that the parameters in the doublet and the triplet are
closely related.
Consequently, we need to consider the constraints from the doublets to determine the superpotential
parameters.
Instead of adjusting the superpotential parameters, then solving the weak doublets and requiring them
to satisfy the low energy data,
we find it is easy for the weak doublets to take their reasonable contents while
the superpotential parameters are determined later.
For those parameters unconstrained by the present data, we simply take
them to be of order one as reasonable inputs.

Although there are several works on fitting the data,
only in \cite{fuku,fits7} the detailed results are presented.
In \cite{fuku} an unacceptably small component of $10_H$ in the MSSM doublet $H_u$ is used
so we will use the numerical results in \cite{fits7}. % on the Yukawa couplings got by fitting the low energy data.
The constraints on the contents of the MSSM doublets give two ratios
\be
r=\frac{\alpha_u^{1}}{\alpha_d^{1}}, ~~~
s=\frac{\alpha_u^{4}}{r\alpha_d^{5}}.
\label{relat}
\ee
In \cite{fits7} the results are given for $tan\beta = 10, 38, 50$ for
$10_{H} + \overline{126}_{H}$ coupling with the matter superfields,
and $tan\beta = 50$ only for $10_{H} + \overline{126}_{H} + 120_{H}$ coupling with the matter superfields.
The former corresponds to taking $Y_{120}=0$ in the later case.

We take $tan\beta = 10$ as an example which corresponds to $r = 13.1538$ and $s = 0.244325 + 0.0495071i$
in \cite{fits7}.
First,
we input the reasonable  contents of the MSSM doublets as
\be
(\alpha_{u}^{1},\alpha_{u}^{4})=(0.8, 0.57)\label{42}
\ee
which, together with the relations (\ref{relat}), give
\be
(\alpha_{d}^{1},\alpha_{d}^{5})
=(0.0608189, 0.170365 - 0.0345208 i).\nonumber
\ee
Then we also use the reasonable  inputs as
\be
(\alpha_{u}^{2},\alpha_{u}^{3})=(0.1,0.1)
\ee
and
\be
(\alpha_{d}^{2},\alpha_{d}^{3},\alpha_{d}^{4})
=( 0.2, 0.5, 0.4).
\ee
From the normalization conditions (\ref{H0}), we have
\be
\alpha_{u}^{5}=0.122882, ~~~\alpha_{d}^{6}=0.718391.\nonumber
\ee

Second,
we require that all the GUT scale masses are of the order of $10^{16}$GeV
except $Q$ which is taken as $\sim 10^{14}$GeV for the seesaw mechanism,
and all the massless couplings are of order 1.
The following massive parameters
\be
\begin{array}{ccccc}
m_{D}=1.0,&m_{1}=0.6,&m_{2}=1.1,&m_{\Phi}=1.2,&E^\prime=0.8,\\
\Phi_{1}=-0.4,&\Phi_{2}=-2,&\Phi_{3}=-1.5,&v_{1}=0.5,&\overline{v}_{2}=0.5\label{massinput}
\end{array}
\ee
are in the unit of $10^{16}$GeV, where in the second line the VEVs are taken larger
values due to normalizations.
Note that in principal the VEVs in (\ref{massinput}) should be determined through
solving the F- and D-flatness conditions after
all the superpotential parameters are fixed first.
The vice verse is also true, since there are extra free parameters in (\ref{Wfull})
not used in the numerical calculations.

Third, for the dimensionless couplings
we take
\be
\lambda_{1}=0.5,~~ \lambda_{2}=0.8, ~~\eta_3=\eta_8=1.0
\ee
and
\be
\eta_{1}=1.4, ~\eta_{2}=1.5, ~\eta_{4}=-0.8, ~\eta_{6}=-1.0.\label{47}
\ee
By solving the linear equations (\ref{eigen}), we get
\bea
\eta_{5}&=&-0.9784+0.00634058 i,  \qquad \eta_{7}=0.892708\, -0.142636 i ,\nonumber \\
\eta_{9}&=& 2.20273 - 0.215545 i,    \qquad \eta_{10}=-2.92267 - 0.283226 i,  \nonumber \\
\eta_{11}&=&1.16329 - 0.120945 i,    \qquad  \eta_{12}=1.28655 + 0.00558605 i, \nonumber \\
\eta_{13}&=&-1.23594 - 0.111137i,   \qquad   \eta_{14}= -0.693669 - 0.3701 i, \nonumber \\
\eta_{15}&=&2.58683\, -0.0386752i. \nonumber
\eea
Putting these parameters into the doublets , we get
\begin{equation}
A_{6\times5}=\scalemath{0.6}
{\left(
\begin{array}{ccccc}
 0.617971\, +0.0555684 i & -2.67484-0.240524 i & -0.854547-0.00371033 i & 0.465314\, -0.0483782 i & 1.23385\, -0.128282 i \\
 0.4892\, -0.00317029 i & 0.7338\, -0.00475544 i & 0.6 & 0.544074\, -0.243254 i & -1.37188+0.0205106 i \\
 0.847319\, -0.00549111 i & 1.12405\, -0.0072845 i & -0.466546 & -1.37188+0.0205106 i & -1.93636+0.0716865 i \\
 1.83712 & -4.20122 & 0.804323\, -0.0787058 i & 0.75 & 1.14887 \\
 0 & -1.06721-0.103419 i & -0.918519 & -0.669531+0.106977 i & 0.520612\, -0.0831829 i \\
 -1.80115 & 1.83712 & 0 & 0.5 & 0.866025 \\
\end{array}
\right)},
\end{equation}
\\
\\
\begin{equation}
B_{4\times5}=\scalemath{0.6}
{\left(
\begin{array}{ccccc}
 2.78438\, +0.0120894 i & 0.6 & 0.919094 & -4.3478 & 0.804323\, -0.0787058 i \\
 0.82093\, +0.0738186 i & 0.7338\, -0.00475544 i & -0.570586+0.00369772 i & -1.06721-0.103419 i & -0.526984 \\
 0.465314\, -0.0483782 i & 0.544074\, -0.243254 i & -1.37188+0.0205106 i & -0.669531+0.106977 i & 0.75 \\
 1.23385\, -0.128282 i & -1.37188+0.0205106 i & -1.93636+0.0716865 i & -1.0256+0.16387 i & -0.583183 \\
\end{array}
\right)}.
\end{equation}
We also get
$\overline{v}_{1}=0.0117188$ through (\ref{v1}) which gives masses to the right handed neutrinos.

Accordingly, for the color triplets, we have
\begin{equation}
A_{6\times6}^{\prime}=\scalemath{0.5}
{\left(
\begin{array}{cccccc}
 -0.617971-0.0555684 i & -1.35872-0.122177 i & -1.51371-0.136114 i & -0.686575-0.00298102 i & 1.00744\, -0.104742 i & 1.89964\, -0.197503 i \\
 0.691833\, -0.00448347 i & 0.599145\, -0.0038828 i & 1.59772\, -0.0103541 i & 0.489898 & 0.279999\, -0.100378 i & -1.58411+0.0236836 i \\
 -0.691833+0.00448347 i & 1.12976\, -0.00732148 i & 0.847319\, -0.00549111 i & -0.92376 & -1.58411+0.0236836 i & -0.189559+0.184499 i \\
 -1.06066 & 1.1 & -2.59808 & 1.60865\, -0.157412 i & 0.612372 & 1.1547 \\
 -1.5 & -2.59808 & -1.23205 & 0 & 1.63299 & 0.866025 \\
 0 & -2.13442-0.206839 i & 0 & 0.6 & -0.54667+0.0873464 i & 1.03081\, -0.164702 i \\
\end{array}
\right)},\label{Anum}
\end{equation}

\begin{equation}
B_{6\times6}^{\prime}=\scalemath{0.5}
{\left(
\begin{array}{cccccc}
 0 & 0 & 0 & 0 & 0 & 0 \\
 1.41436\, +0.00614096 i & 0.489898 & 0.92376 & 0.6 & -2.42487 & 1.60865\, -0.157412 i \\
 1.5757\, +0.00684149 i & 1.30639 & 0.69282 & -2.42487 & -1.57658 & 0 \\
 0.659566\, +0.0593086 i & 0.599145\, -0.0038828 i & -1.12976+0.00732148 i & -2.13442-0.206839 i & 0 & 1.1 \\
 1.34325\, -0.139656 i & 0.279999\, -0.100378 i & -1.58411+0.0236836 i & -0.54667+0.0873464 i & -1.45779+0.232924 i & 0.612372 \\
 1.89964\, -0.197503 i & -1.58411+0.0236836 i & -0.189559+0.184499 i & -1.03081+0.164702 i & -0.773108+0.123526 i & -1.1547 \\
\end{array}
\right)},\label{Bnum}
\end{equation}
and
\begin{equation}
C_{6\times6}^{\prime}=\scalemath{0.6}
{\left(
\begin{array}{cccccc}
 -0.5684 & -1.06066 & -1.5 & 0 & 0.707107 & -0.707107 \\
 -0.0232019 & 0.01 & 0 & 0 & 0 & 0 \\
 -0.0328125 & 0 & 0.01 & 0 & 0 & 0 \\
 0 & 0 & 0 & 0.01 & 0 & 0 \\
 -0.0147947+0.00236388 i & 0 & 0 & 0 & 0.01 & 0 \\
 0.0147947\, -0.00236388 i & 0 & 0 & 0 & 0 & 0.01 \\
\end{array}
\right)}.\label{Cnum}
\end{equation}

In choosing the numerical inputs above, we do not
fine-tune any number besides solving the linear equations (\ref{eigen}).
instead, we have imposed the  constraints on the inputs that only few
small or large matrix elements can exist and all the eigenvalues of the doublet and the triplet mass matriices
are of order $\Lambda_{GUT}$.
%However, we have imposed very strong constraints on the inputs so that no matrix element
%bigger than 3.0 is allowed, and the few small elements must not induce
%small eigenvalues.
Then, no large splitting exists in the spectrum so that
the GUT scale threshold effects are small.
Consequently,
the predictions on proton decay in the following are not tuned
which will be taken as natural estimations in the model \cite{cz2}.

\section{The proton decay via dimension five operator}

In SUSY GUT models, proton decays are dominated by the  baryon and lepton number violating
operators of dimension-five
\begin{equation}
W=C_{L}^{ijkl}Q^{i}Q^{j}Q^{k}L^{l}
\end{equation}
dressed mainly by the wino components of the charginos\cite{su5pd4}.
The coefficients are\cite{so10d,fuku}
\begin{equation}
C_{L}^{ijkl}=\left(
               \begin{array}{ccc}
                 Y_{10}^{ij} & Y_{126}^{ij} & 0 \\
               \end{array}
             \right)\left(
                      \begin{array}{c}
                        M_{C}^{-1} \\
                      \end{array}
                    \right)\left(
                             \begin{array}{c}
                               Y_{10}^{kl} \\
                               Y_{126}^{kl} \\
                               \sqrt{2}Y_{120}^{kl} \\
                             \end{array}
                           \right).\label{CL}
\end{equation}
Here $M_C$ is effective triplet mass matrix got by integrating out those fields which do not couple
with $Q$ or $L$ in  (\ref{massTf2},\ref{Anum}-\ref{Cnum}).
For  $tan\beta=10$, we have
\be
(M_{T}^{eff})_{33}=\left(
\begin{array}{ccc}
 \infty & -62.1295-90.5827 i & -324.779+31.8573 i \\
 164.562\, -5.04775 i & -400.812-7.02252 i & 133.955\, -12.2758 i \\
 -200.267-73.1462 i & 411.089\, +161.251 i & 525.038\, +62.3212 i \\
\end{array}
\right).
\ee
It is obvious that proton decays are dominated by the contributions from $\overline{126}$ and/or $120$.
Then the decay rates are proportional to $|Y_{126}|^4$ or $|Y_{126}Y_{120}|^2$.
These Yukawa couplings extracted from \cite{fits7} are
\begin{eqnarray}
Y_{10}=\left(
\scalemath{0.7}
{
\begin{array}{ccc}
 0.000180154\, +0.000194604 i & 0.000928682\, +0.000572057 i & 0.000667056\, -0.000681864 i \\
 0.000928682\, +0.000572057 i & 0.00559167\, +0.0000902319 i & -0.00390991-0.0113164 i \\
 0.000667056\, -0.000681864 i & -0.00390991-0.0113164 i & 1.02374 + 5.28625 \times 10^{-7}i\\
\end{array}
}
\right)\nonumber
\end{eqnarray}
and
\begin{eqnarray}
Y_{126}=\left(
\scalemath{0.7}
{
\begin{array}{ccc}
 -0.000246286-0.000272057 i & -0.00130341-0.000802886 i & -0.000936218+0.000957003 i \\
 -0.00130341-0.000802886 i & -0.00607135-0.00273387 i & 0.00548759\, +0.0158826 i \\
 -0.000936218+0.000957003 i & 0.00548759\, +0.0158826 i & 0.00938846\, -0.0581996 i \\
\end{array}
}
\right).
\end{eqnarray}

The dominant proton decay mode via the dimension five operator with the Wino dressing diagram is
$p\to K^{+}\overline{\nu}$.
The decay rate  is approximately
\begin{eqnarray}
\Gamma(p \to K^{+} \bar{\nu}) &\simeq & \Gamma(p \to K^{+} \bar{\nu_{\tau}})\nonumber\\
                               &=&   \frac{m_{p}}{32\pi f_{\pi}^{2}}|\beta_{H}|^{2}\times |A_{L} A_{S}|^{2}\times(\frac{\alpha_{2}}{4\pi})^{2}\frac{1}{m_{S}^{2}} |C_{L}^{1123}-C_{L}^{1213}+\lambda(C_{L}^{1223}-C_{L}^{2213})|^{2}\nonumber\\
                            &&\times  5.0 \times 10^{31}[{\rm yrs}^{-1}/{\rm GeV}],\label{pdform}
\end{eqnarray}
which gives the partial lifetime
\begin{eqnarray}
\tau(p\to K^{+}\overline{\nu})= 3.88362\times10^{34}{\rm yrs}
\end{eqnarray}
for $tan\beta = 10$. In (\ref{pdform}) we have used the MSSM and hadronic parameters taken from \cite{fuku}.
The above numerical predictions depend on the inputs (\ref{42}-\ref{47}).
%Among them, the most meaningful parameter is the component $\alpha_u^4$ whose
%range is not well determined in the literature.
%When we choose a smaller $\alpha_{u}^{4} = 0.3 $, and change
%$\phi_{3}$ in the range $(12.1 \sim 14.1)$ in units of $10^{16}$GeV,
We have changed several VEVs by a factor of two around
$2 \times 10^{16}$GeV, and find that
the proton partial lifetime varies in
$10^{34} \sim 10^{36}$yrs for $\tan\beta=10$, consistent with the present data.

We also calculate the proton decay  partial lifetimes with constraints given in \cite{fits7}
which are
\begin{eqnarray}
\tau(p\to K^{+}\overline{\nu})= 5.52536\times10^{34}{\rm yrs}
\end{eqnarray}
for $tan\beta = 38$ and
\begin{eqnarray}
\tau(p\to K^{+}\overline{\nu})= 6.85908 \times10^{33}{\rm yrs}
\end{eqnarray}
for $tan\beta = 50$ with $10_{H} + \overline{126}_{H}$ giving fermion masses.
As we can see, the proton decay can be suppressed even for large $tan\beta$ and there
is no obvious $tan\beta$ dependence in the partial lifetimes
which are all consistent with the present lower limit $6.6\times 10^{33}$ years\cite{pddata}.

However, using the results by fitting the data with $10_{H} + \overline{126} + 120_{H}$\cite{fits7},
\begin{eqnarray}
\tau(p\to K^{+}\overline{\nu})= 3.59502\times10^{30}{\rm yrs},
\end{eqnarray}
which is much lower than the data\cite{pddata}.
This can be tracked back to the Yukwawa couplings given in \cite{fits7}.
Compared to the fitting without $120_H$, %in the present case
all the  entries in the  Yukawa couplings $Y_{126}$ and  $Y_{120}$ appearing in (\ref{CL},\ref{pdform})
are lager by one order of magnitude.
Thus the results in this case is very difficult to understand,
since without  $120$ the fitting is rather good except small values like $m_e$\cite{fits3},
thus $120$ is probably playing a minor role in the fitting.
Also, the top quark mass calculated using the results  in \cite{fits7}
is generally larger than the input used by the same paper,
which exhibits probably numerical inconsistency in \cite{fits7}.

There are also sub-dominant decay modes whose partial lifetimes are also calculated. The results
are summarized in Table 3. Again, there are conflicts when $120_H$ is included to contribute to the fermion masses.

\begin{table}
\begin{center}
\begin{tabular}{|c|c|c|c|c|c|}
  %\hline
  \hline
  \multicolumn{1}{|c|}{}&
  \multicolumn{1}{|c|}{Lower limit}&
  \multicolumn{3}{|c|}{$10_{H} + \overline{126}_{H}$}&
  \multicolumn{1}{|c|}{$10_{H} + \overline{126}_{H} + 120_{H}$}\\
  \hline
  Decay mode      &  \cite{pddata}   & $tan\beta = 10$ & $tan\beta = 38$ & $tan\beta = 50$ & $tan\beta = 50$\\
  \hline
  $p\to K^{+}\overline{\nu}$ & $6.6\times 10^{33}$& $3.88\times10^{34}$  & $5.53\times10^{34}$  & $6.86\times10^{33}$ & $3.60\times10^{30}$\\
  \hline
  $p\to K^{0}e^{+}$  & -  &  $2.67\times10^{39}$  &  $3.42\times10^{38}$  &  $3.83\times10^{37}$  & $6.96\times10^{35}$\\
  \hline
  $p\to K^{0}\mu^{+}$  &$6.6\times 10^{33}$  &  $1.01\times10^{36}$      &$1.35\times10^{35}$        &  $1.60\times10^{34}$  & $2.69\times10^{32}$\\
  \hline
  $p\to \pi^{+}\overline{\nu}$  &$3.9\times 10^{32}$   &  $2.55\times10^{37}$  & $2.85\times10^{36}$ & $2.67\times10^{35}$   & $2.71\times10^{32}$\\
  \hline
  $p\to \pi^{0}e^{+}$   &$1.7\times 10^{34}$  &  $4.64\times10^{40}$   & $6.16\times10^{39}$    & $7.33\times10^{38}$   & $2.64\times10^{36}$\\
  \hline
  $p\to \pi^{0}\mu^{+}$   &$7.8\times 10^{33}$  &  $1.75\times10^{37}$    & $2.43\times10^{36}$     & $3.06\times10^{35}$    & $1.02\times10^{33}$\\
  \hline
\end{tabular}
\end{center}
\caption{Partial lifetimes of proton decay in years using different Higgs to fit fermion masses.
Dimension five operators with charged wino dressing are used only.
Inputs (\ref{42}-\ref{47}) are used.
The Yukawa couplings  are taken from \cite{fits7}. }
\end{table}

\section{Summary}

In this article we have examined the  renormalizable SUSY SO(10) model \cite{cz2}.
Without any fine-tuning of the parameters,
we have shown how to construct MSSM doublets, to determine
the parameters of the model,
and to predict on proton decay rates.
We find that in the case using with $10_{H} + \overline{126}_{H}$ to fit fermion masses and mixing,
proton decay lifetimes are consistent with the experiment.
In the case using also $120_{H}$ to fit the data,
proton decay too fast.
However, we find the numerical results with  $120_{H}$ may not be consistent,
and independent check of the same study is highly called for.

As in all renormalizable SUSY GUT models,
the large representations used in \cite{cz2}
contribute largely to the $\beta$-function of the GUT gauge coupling.
Then
the GUT gauge coupling blows up quickly above the unification scale and cause the
non-perturbative problem.
However, the universe was in the GUT symmetric phase at very high temperature in its very early stage.
There happened a phase transition  and the GUT symmetry was broken
when the universe was cooling down.
However, this phase transition has only been well
studied in very simple models in the perturbative region.
Without definite conclusions on  the phase transition
especially in the models in non-perturbative region,
the running behavior of the GUT gauge coupling before this phase transition  may not be
a real difficulty.

\end{document}